\documentclass[aps,pra,twocolumn,amsfonts,amssymb,amsmath,showpacs,showkeys,
floatfix,nofootinbib,groupedaddress,superscriptaddress,citesort]{revtex4}

\usepackage{graphicx}\usepackage{pdfpages}\usepackage{palatino}
\usepackage{tikz} \usepackage{amsmath}\usepackage{mathrsfs}
\usepackage{amsfonts}
\usepackage{amstext}
\usepackage{amsmath}
\usepackage{amssymb}\usepackage{mathrsfs}
\usetikzlibrary{shapes.geometric, arrows}
\usepackage{booktabs}
\usepackage{multirow}
\usepackage{threeparttable}

\newtheorem{definition}{Definition}
\def\ra{\rangle}

\begin{document}
\title
{Network mechanism for generating genuinely correlative Gaussian states }
\thanks{ The paper is dedicated to Prof. Jinchuan Hou on the occasion of his 70th birthday.}

\author{Zhaofang Bai}
\email{baizhaofang@xmu.edu.cn} \address{School of Mathematical
Sciences, Xiamen University, Xiamen, Fujian, 361000, China}

\author{Shuanping Du}
\thanks{Corresponding author}
\email{dushuanping@xmu.edu.cn} \address{School of Mathematical
Sciences, Xiamen University, Xiamen, Fujian, 361000, China}

\begin{abstract}

Generating a long-distance quantum state with genuine quantum correlation (GQC) is one of the
most essential functions of quantum networks to support quantum communication. Here, we pro-
vide a deterministic scheme for generating multimode Gaussian states with certain GQC (including
genuine entanglement). Efficient algorithms of generating multimode states are also proposed.
Our scheme is useful for resolving the bottleneck in generating some multimode Gaussian states
and may pave the way towards real world applications of
preparing multipartite quantum states in current quantum technologies.


\end{abstract}
\maketitle

\section{  Introduction}

The existence of multipartite quantum states that cannot
be prepared locally is at the heart of many communication
protocols in quantum information science, including quantum teleportation \cite{Bennett4}, dense coding \cite{Bennett5}, entanglement-based quantum key distribution \cite{Scarani}, and the violation of Bell inequalities \cite{Bell,Brunner}. Therefore, preparing a desired multipartite quantum
state from some available resource states under certain quantum operations is of great foundational and practical interest.
Among the quantum correlation,  the entanglement is used firstly as a physical
resource, so preparing bipartite entangled states under the class of local operations and classical communication have been studied extensively
\cite{Bennett1, Bennett2, Bennett3, Horodecki1, Beigi}. However, recent study  has  undergone a major development to multipartite scenarios featuring several independent sources that each distributes a resource state \cite{Luo}. The independence of sources reflects a network structure
over which parties are connected.  This is not only due to researcher's interests in understanding quantum theory
and its relationship in more sophisticated and qualitative scenarios \cite{Acin,Raussendorf,Walther,Briegel,Halder} but also technological
developments towards scalable quantum networks \cite{Luo,Kimble,Wehner,Kozlowski}.

Quantum networks are of high interest nowadays, which are the way how quantum sources distribute particles to different parties in the network. Quantum networks
play a fundamental role in the long-distance secure communication \cite{Poppe, Hammerer}, exponential gains in communication complexity \cite{Guerin}, clock synchronization \cite{Komar} and distributed quantum computing \cite{Cirac}.
Most importantly, for the last two
decades, generating a multipartite state via appropriate quantum operations from states having lesser number of parties with the assurance of multipartite correlation has been regarded as a benchmark in the development of quantum networking test beds \cite{Sang, Navascues, Jones, Prit}.
The network mechanism has been used to generate special multipartite states which play an important role for quantum computation and quantum communication tasks \cite{Pirk1,Pirk2,Gyong,Migue,Azuma,Migue2}.

In this research direction, the infinite dimensional counterpart of the above-mentioned state preparation method should
be explored. In particular, Gaussian states constitute a wide
and important class of quantum states, which serve as the
basis for various types of continuous-variable quantum information processing \cite{Weedbrook}. The goal of this paper is to  find the Gaussian networks \cite{Ghalaii} mechanism for generating multimode Gaussian states.


We provide a protocol for generating multimode Gaussian  states with certain amount of genuine Gaussian  quantum correlation (GGQC) over a large quantum Gaussian network. This provides a generic method to deterministically generate  multimode Gaussian states with GQC.

Precisely, we consider quantum Gaussian networks in continuous-variable (CV) systems consisting of spatially separated nodes (parties) $P_1, P_2, \ldots , P_N$, $s$ ($s\leq N$) independent sources, each generating an $n_i$-mode Gaussian state $|\phi_l\rangle$ $(l=1,2,\ldots,s)$. And each  node $ P_i$ consists of $m_i$ modes \cite{Luo}. If the nodes share more than one source with other nodes, we call them intermediate nodes. Other nodes are called extremal nodes. Our protocol is to apply $2-$mode Gaussian unitary operations $U_i$ at intermediate parties and the 2 modes are from different sources. Define Gaussian operation $$\Phi_i(\cdot)=U_i\otimes I_{\overline{i}}\cdot U_i^{\dag}\otimes I_{\overline{i}} $$ and $\Phi=\Pi_i \Phi_i$, where $I_{\overline{i}}$ denotes the identity operator acting on
the rest of the modes except modes acted by $U_i$ (see Figure 1). We examine the relation between  GGQC of resultant state $\Phi(\bigotimes_l |\phi_l\rangle)$  and GGQC of the source states $\{|\phi_l\rangle, l=1,2, \cdots, s\} $. And show that to make a quantum network having certain amount of  GGQC,  one needs to create  source states containing at least the same amount of GGQC, since the minimum GGQC among the source states coincides with the GGQC of the resultant state $\Phi(\bigotimes_l |\phi_l\rangle)$, obtained after applying optimal Gaussian unitary operations on the initial state $\bigotimes_l |\phi_l\rangle$. We note that all Gaussian unitary operations that maximize the GGQC of source states in our scheme are called optimal Gaussian unitary operations. 

The paper is organized as follows. After reviewing detailed definitions and notations of  continuous-variable  systems in Sec. II. We provide a GGQC measure in Sec. III. We then give our protocol for generating  multimode Gaussian  states with certain amount of GGQC in Sec. IV. The last section is a summary of our findings. The Appendix gives the proof of our results.

\begin{figure}[htbp]
\centering
\includegraphics[scale=0.35]{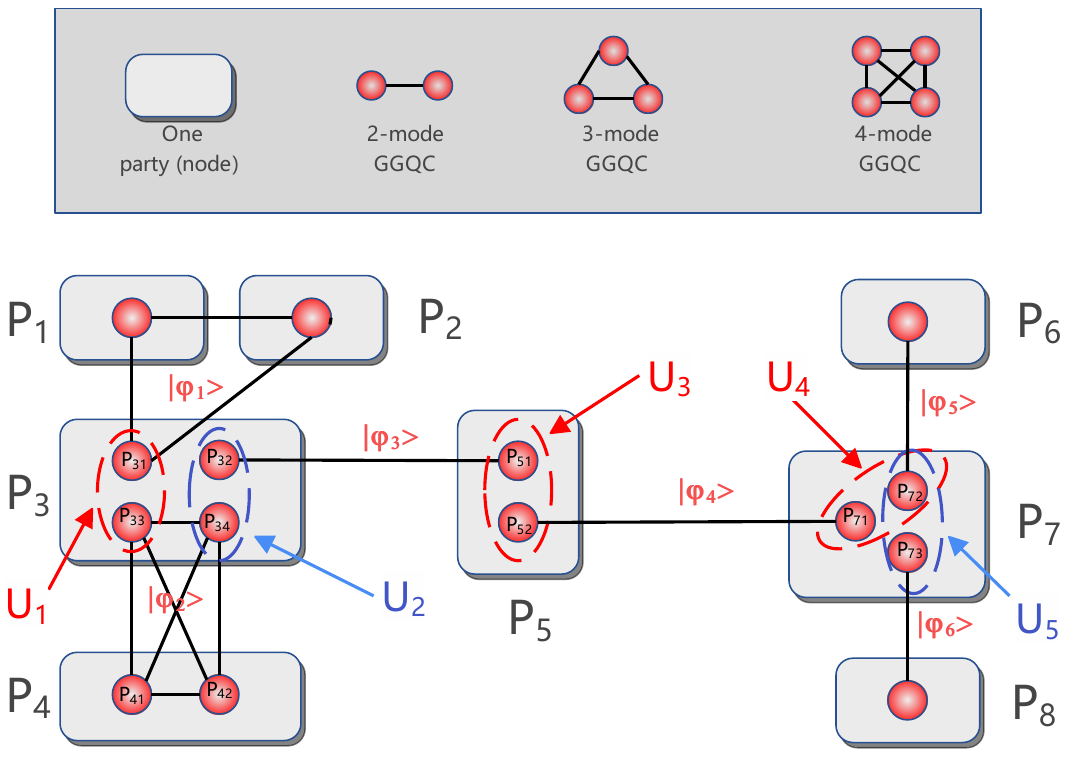}
\caption{ Schematic representation of creating multimode Gaussian states by applying
arbitrary 2-mode Gaussian unitary operation at intermediate parties. Here each ball denotes one mode, $N=8$, $s=6$, $m_{1,2,6,8}=1$, $m_{4,5}=2$, $m_3=4$, $m_7=3$.
The initial state is $\otimes _{l=1}^6|\phi_l\rangle$, $2-$mode Gaussian unitary operation $U_1$ acts on $P_{31}$ and $P_{33}$ (contained in intermediate node $P_3$), $P_{31}$ coming from $|\phi_1\rangle$ and  $P_{33}$ from $|\phi_2\rangle$, respectively. Any other $U_i (i=2,\cdots, 5)$ acts  two modes which  is from $|\phi_i\rangle$, $|\phi_{i+1}\rangle$. 
 We are aimed  to find out the optimal
Gaussian unitary operations $\{U_i\}$ such that the resulting multimode state
possess maximal GGQC.}

\end{figure}

\section{Background on Gaussian systems}

We now review some definitions and notations concerning
 Gaussian quantum information theory (\cite{Weedbrook,Sera,Hou}).
Recall that an $n$-mode Gaussian system  is  determined by $2n$-tuple
$\widehat{R}=(\hat{Q}_1,\hat{P}_1,\cdots,\hat{Q}_n,\hat{P}_n)$ of self-adjoint
operators with state space $H=H_1\otimes H_2\otimes\cdots\otimes
H_n$, where $\hat{P}_r,\hat{Q}_r$ are respectively
 the position and momentum operators of the $r$th-mode which act on the separable infinite
dimensional complex Hilbert space $H_r$. As it is well known,
$\hat{Q_r}=(\hat{a_r}+\hat{a_r}^t)/\sqrt{2}$ and
$\hat{P_r}=-i(\hat{a_r}-\hat{a_r}^t)/\sqrt{2}$
($r=1,2,\cdots,n$) with
 $\hat{a}_r^t$ and
$\hat{a}_r$ being the creation and annihilation operators in the $r$th
mode $H_r$, which satisfy the canonical commutation relation (CCR)
$$[\hat{a}_r,\hat{a}_s^t]=\delta_{rs}I\ {\rm and}
\ [\hat{a}_r^t,\hat{a}_s^t]=[\hat{a}_r,\hat{a}_s]=0,\ \
r,s=1,2,\cdots,n.$$ Denote by ${\mathcal S}(H)$ the set of all
quantum states in a system described by $H$ (the positive operators on
$H$ with trace 1). The
characteristic function $\chi_{\rho}$ for  any state $\rho\in{\mathcal S}(H)$ is defined as
$$\chi_{\rho}(z)={\rm tr}(\rho W(z)),$$ where
$z=(x_{1}, y_{1}, \cdots, x_{n}, y_{n})^{\rm T}\in{\mathbb R}^{2n}$,
$ W(z)=\exp(i\widehat{R}\Omega z)$ is the Weyl displacement operator, $\Omega=\oplus \left(\begin{array}{cc}0&1\\-1&0\end{array}\right)$.
Let ${\mathcal {FS}}(H)$ be the set of all quantum states with
finite second moments, that is, $\rho\in{\mathcal {FS}}(H)$ if ${\rm
Tr}(\rho \hat{R_r})<\infty$ and ${\rm Tr}(\rho \hat{R_r}^2)<\infty$
for all $r=1,2,\ldots, 2n$. For $\rho\in{\mathcal {FS}}(H)$, its
first moment vector
$$\begin{array}{ll}{\mathbf d}_\rho &=(\langle\hat R_1 \rangle, \langle\hat R_2
\rangle, \ldots ,\langle\hat R_{2n} \rangle)^{\rm T}\\
&=({\rm tr}(\rho
\hat R_1), {\rm tr}(\rho \hat R_2), \ldots, {\rm tr}(\rho \hat
R_{2n}))^{\rm T}\in{\mathbb R}^{2n}\end{array}$$ and its second moment matrix
$$\Gamma_\rho=(\gamma_{kl})\in M_{2n}(\mathbb R)$$
defined by $\gamma_{kl}={\rm tr}[\rho
(\Delta\hat{R}_k\Delta\hat{R}_l+\Delta\hat{R}_l\Delta\hat{R}_k)]$
with $\Delta\hat{R}_k=\hat{R}_k-\langle\hat{R}_k\rangle$
(\cite{Braunstein}) are called the mean and the covariance matrix (CM) of
$\rho$ respectively.  Here
$M_k(\mathbb R)$ stands for the algebra of all $k\times k$ matrices
over the real field $\mathbb R$. Note that a CM $\Gamma$ must be real symmetric and satisfy the uncertainty
condition $\Gamma +i\Omega\geq 0$.
A Gaussian state $\rho\in{\mathcal {FS}}(H)$ is such a state of which the  characteristic function
$\chi_{\rho}(z)$ is of the form
\begin{eqnarray*}\chi_{\rho}(z)=\exp[-\frac{1}{4}z^{\rm T}\Gamma_\rho z+i{\mathbf
d}_\rho^{\rm T}z].
\end{eqnarray*}
For an $n$-mode CV system determined by ${R}=(\hat
R_1,\hat R_2,\cdots,\hat
R_{2n})=(\hat{Q}_1,\hat{P}_1,\cdots,\hat{Q}_n,\hat{P}_n)$,  it is known that a unitary operation $U$ is Gaussian if and only if
there is a vector $\mathbf m$   in ${\mathbb R}^{2n}$ and a   matrix
$\mathbf S \in {\rm Sp}(2n,\mathbb R)$ such that $U^{\dag}R^tU = \mathbf SR^t +\mathbf m$ (\cite{Weedbrook}), where
${\rm Sp}(2n,\mathbb R)$ is
the symplectic group consisting of all
$2n\times 2n$ real matrices $\bf S$ that satisfy
$\mathbf S\in{\rm Sp}(2n,\mathbb R)\Leftrightarrow \mathbf S\Omega \mathbf S^{\rm
T}=\Omega$.
Thus, every Gaussian unitary operation $U$ is determined
by some affine symplectic map $(\mathbf S, \mathbf m)$ acting on the
phase space, and can be parameterized as $U=U_{\mathbf S, \mathbf m}$. It follows that, if $U_{\mathbf S,\mathbf m}$ is a Gaussian unitary
operation, then, for  any $n$-mode
 state $\rho$   with CM $\Gamma_\rho$ and mean $\mathbf d_\rho$,
the  state $\sigma=U_{\mathbf S,\mathbf m}\rho U_{\mathbf
S,\mathbf m}^{\dag}$ has the CM $\Gamma_{\sigma} = \mathbf S\Gamma_{\rho}
\mathbf S^{\rm T}$ and the mean $\mathbf d_{\sigma} =\mathbf m + \mathbf
S\mathbf d_{\rho}$.

\section{A GGQC measure}

An amazing feature of quantum mechanics is the existence of quantum correlations. Various methods for quantifying quantum correlations
are one of the most actively researched subjects in the past few decades \cite{Horodecki1, Weedbrook,Modi}.
Measurements of quantum correlations have
played an important role in understanding the properties of quantum many-body systems and their non-classical behaviors.

In the following, we will propose a definition of GGQC measure.
To the best of our knowledge, this is the first thought to define
 multimode genuine Gaussian quantum correlation measure beyond entanglement.
In addition,  a pure Gaussian state with genuine Gaussian quantum correlation under our GGQC measure is also genuine entanglement \cite{Navascues,Sen2}.

For any $n-$mode Gaussian state $\rho_{A_1,A_2,\ldots, A_n}$ on $(H_{A_1}\otimes
H_{A_2}\otimes\cdots \otimes H_{A_n})$, its CM can be
represented as
\begin{equation}\Gamma_{\rho_{A_1,A_2,\ldots, A_n}}=\left(\begin{array}{cccc} A_{11} & A_{12} &\cdots & A_{1n}\\
A_{21} & A_{22} &\cdots & A_{2n} \\ \vdots & \vdots & \ddots &\vdots
\\ A_{n1} & A_{n2} & \cdots & A_{nn}  \end{array}\right),\end{equation}
where $A_{jj}\in M_{2}(\mathbb R)$ is the CM of the reduced state
$\rho_{A_j}={\rm Tr}_{A_j^c}(\rho_{A_1,A_2,\ldots, A_n})$ of
$\rho_{A_1,A_2,\ldots, A_n}$,  $A_j^c=\{A_1,\ldots, A_{j-1},
A_{j+1},\ldots, A_n\}$, namely, $A_{jj}=\Gamma_{\rho_{A_j}}$, and off-diagonal blocks
$A_{ij}\in M_{2}(\mathbb R)$ encode the intermodal correlations  between
subsystems ${A_i}$ and $A_j$.
For any $(n_1+n_2)$-mode $2$-partite state $\rho$ with CM
$$\Gamma_{\rho}=\left(\begin{array}{cc}A&C\\C^t&B\end{array}\right),$$
the quantity $${\mathcal M}(\rho)=1-\frac{\det(\Gamma)}{\det(A)\det(B)}$$ is discussed in  \cite{Castro,Liu,Hou}. It is evident that, for any $2$-partition ${\mathcal P}$ of $n$-mode system $A_1A_2\ldots A_n$,
there exists  a permutation $\pi$ of $(1,2,\ldots
n)$ and positive integers $n_1,n_2$ with
$n_1+n_2=n$ such that
$${\mathcal P}
=   A_{\pi(1)}\ldots A_{\pi(n_1)}|A_{\pi(n_1+1)}\ldots
A_{\pi(n)}.$$
One can compute the ${\mathcal M(\rho)}$ with respect to $\mathcal P$ denoted by ${\mathcal M}_{\rho}( \mathcal P )$. Now we provide the definition of our GGQC measure.
\begin{definition}For any $n$-mode Gausian state $\rho$, define the  quantity $\mathcal{ GM}(\rho)=\min_{{\mathcal P}} {\mathcal M}_{\rho}(\mathcal P)$, here ${\mathcal P}$ runs over all 2-partitions.
\end{definition}
Note that any 2-partition $\mathcal P$ corresponds a subset $\alpha$ of $\{1,\ldots, n\}$. Let $\mathcal D_{\rho}({\alpha})$ be the principle minor that lies in the rows and columns  of $\Gamma_{\rho} $ indexed by $\alpha$ and $\overline{\alpha}$ denotes its complement set. Then ${\mathcal M}_{\rho}( \mathcal P )$ is also written as ${\mathcal M}_{\rho}( \alpha )$ and
\begin{equation}\mathcal{ GM}(\rho)=\min_{\alpha}\{ 1-\frac {\det(\Gamma_{\rho})}{\mathcal D_{\rho}({\alpha})\mathcal D_{\rho}({\overline{\alpha} })}\}.
\end{equation}

In fact, $\mathcal{GM}$
has the following  properties which satisfy the basics of Gaussian quantum
correlation measure \cite{Modi,Castro, Liu, Hou, Ciccaarello,Girolami}.

(1) $0\leq \mathcal {GM}(\rho)\leq 1$.

(2) $\mathcal {GM}(\rho)=0$ if and only if $\rho$ is a product state with respect to at least one modal bipartition.

(3) $\mathcal {GM}$  is invariant under any permutation of
system, that is, for any permutation $\pi$ of $(1,2,\ldots, n)$,
denoting by $\rho_{A_{\pi(1)},A_{\pi(2)},\ldots, A_{\pi(n)}}$ the
state obtained from the state $\rho_{A_1,A_2,\ldots, A_n}$ by
changing the order of the subsystems according to the permutation
$\pi$, we have
$$\mathcal {GM}(\rho_{A_{\pi(1)},A_{\pi(2)},\ldots, A_{\pi(n)}})=\mathcal {GM}(\rho_{A_1,A_2,\ldots,
A_n}).
$$

(4) $\mathcal {GM}$ is invariant under  locally
Gaussian unitary operations on $H_{A_1}\otimes
H_{A_2}\otimes\cdots \otimes H_{A_n}$.

(5) $\mathcal {GM}$ is nonincreasing under local Gaussian operations.

It is evident that if $\mathcal {GM}(|\phi\rangle)\neq 0$, then $|\phi\rangle$ is not a product state with respect to any 2-partition of $\{1,2,\cdots, n\}$,
so we say $|\phi\rangle$ is  genuinely correlative. The property is harmonic with the key generalized geometric measure of genuine entanglement which is defined as the shortest distance of a given multimode state from a nongenuinely multimode entangled state \cite{Sen2}. This implies $|\phi\rangle$ is genuinely correlative if and only if $|\phi\rangle$ is  genuinely entangled. Genuine correlation and genuine entanglement \cite{Navascues,Sen2} are not coincident for mixed states since
$\mathcal {GM}(\rho)\neq 0$ if and only if $\rho$ is not a product state with respect to any 2-partition of $\{1,2,\cdots, n\}$.
Compared with some known entanglement measures, such as  the distillable entanglement, the entanglement of formation, the entropy of entanglement and the generalized geometric measure \cite{Weedbrook, Sen2}, $\mathcal {GM}$ is more easy to calculate since all 2-partitions of $\{1,2,\cdots, n\}$ are finite and no optimization process is involved. In the next paragraph, we will compute the value of $\mathcal {GM}$ for some important Gaussian states. To the best of our knowledge, $\mathcal {GM}$ is the only known  multimode genuine Gaussian quantum correlation measure beyond entanglement. Since genuine multipartite entanglement has  become a
standard for quantum many-body experiments \cite{Lu, Gross,Yao,Wang},  $\mathcal {GM}$ may become one of the best prospects for unveiling essential Gaussian quantum correlation of multimode systems.

For any 2-mode Gaussian pure state $|\phi\rangle$, under some suitable local Gaussian unitary
operation, its CM can be reduced to the standard form \cite{Lami2}
$$\Gamma_{|\phi\rangle}=\left(\begin{array}{cc}
                \gamma I_2  & \sqrt{\gamma^2-1}C\\
                \sqrt{\gamma^2-1}C & \gamma I_2\end{array}\right), \quad
C=\text{diag}(1,-1),$$ $\gamma\geq 1$ is the single-mode mixedness factor and $I_2$ is the $2\times 2$ unit matrix. A direct computation shows $$\mathcal{
GM}(|\phi\rangle)={\mathcal
M}(|\phi\rangle)=1-\frac 1{\gamma^4}.$$ In fact, using the standard form of CM for any 2-mode Gaussian state $\rho$,

 $$\Gamma_{\rho}=\left(\begin{array}{cc}
                aI_2 & C\\
                C & bI_2\end{array}\right), \quad
C=\left(\begin{array}{cc}
                c & 0\\
                0 & d \end{array}\right),$$ $a\geq 1, b\geq 1, c,d\in{\mathbb R}$ \cite{Duan1,Simon1}, one can obtain $$\mathcal{
GM}(\rho)=1-\frac{(ab-c^2)(ab-d^2)}{a^2b^2}.$$

For the case of 3-mode, we analyse a pure state $|\phi_{\gamma}\ra$ prepared by
combining three single-mode squeezed states in a tritter (a
three-mode generalization of a beam splitter), which
possesses the CM, given by \cite{Fer},
\begin{equation}\left(\begin{array}{cccccc}
\mathcal{R}_+ &0        &\mathcal{S} &0  &\mathcal{S} &0 \\
0             &\mathcal{R}_-&0& \mathcal{-S} &0  &\mathcal{-S}  \\
\mathcal{S}   & 0 &\mathcal{R}_+ &0  &\mathcal{S} &0 \\
0             &\mathcal{-S}&0&\mathcal{R}_- &0  &\mathcal{-S} \\
\mathcal{S}   & 0 &\mathcal{S} &0  &\mathcal{R}_+ &0 \\
0             &\mathcal{-S}&0&\mathcal{-S} &0  & \mathcal{R}_-
\end{array}\right),\end{equation}
where $\mathcal{R}_{\pm}=\cosh(2\gamma)\pm \frac 13 \sinh(2\gamma)$ and $\mathcal{S}=-\frac 23\sinh(2\gamma)$. By a direct computation,
$$\mathcal{GM}(|\phi_{\gamma}\ra)=1-\frac1 {{\mathcal R}_+^2{\mathcal R}_-^2}=1-\frac {81}{(5+4\cosh(4\gamma))^2}.$$
Therefore we provide a formula of $\mathcal{GM}$ as a function of the squeezing strength $\gamma$. It is evident that the $\mathcal{GM}$ approaches its maximum value 1 as $\gamma\rightarrow \infty$. Combining this and computing formula of the generalized geometric  entanglement measure ${\mathcal G}(.)$ on $|\phi_{\gamma}\ra$\cite{Sen2}, we can find an interesting fact  $$\mathcal{GM}(|\phi_{\gamma_1}\ra)\leq \mathcal{GM}(|\phi_{\gamma_2}\ra)\Leftrightarrow\mathcal{G}(|\phi_{\gamma_1}\ra)\leq \mathcal{G}(|\phi_{\gamma_2}\ra)$$ for pure states $|\phi_{\gamma_1}\ra, |\phi_{\gamma_2}\ra$. This tells that the measures $\mathcal{GM}$ and $\mathcal{G}$  have
the same order on three single-mode squeezed states in a tritter.

\section{ Generation of multimode Gaussian states with GGQC}

We now introduce a procedure for preparing a Gaussian network to be in a large multimode state  with certain amount of GGQC.
Let us consider a Gaussian network with $N$ parties (nodes) $P_1, P_2, \ldots , P_N$, $s$ ($s\leq N$) independent sources, each generating an $n_i$-mode Gaussian state $|\phi_i\rangle$ $(i=1,2\ldots,s)$. Then the quantum Gaussian network is a system involving $n=\sum_{i=1}^s
 n_i$ modes, the initial state is given by $\rho=\otimes |\phi_i\rangle$. Our main result reads as follows. 

{\bf Theorem 4.1.} For initial state $\rho=\otimes |\phi_i\rangle$, there exist optimal Gaussian unitary operations such that the resultant states  give maximal GGQC by $$\max_{\{U_i\}}\mathcal{ GM}((\Phi(\rho))=\min_i \{\mathcal{ GM}(|\phi\ra_i)\}.$$

Let us now stress some key points about Theorem 4.1.

(i) Theorem 4.1 provides an explicit formula for the maximum GGQC that can
be generated by our protocol.  We need to prepare a number of low mode source states containing at least the same amount of GGQC in order to create a multimode Gaussian state with certain amount of GGQC. 
Note that the property of genuine correlation and genuine entanglement \cite{Sen2} is harmonic for any pure Gaussian state,  our protocol also supports generation of multipartite genuinely entangled states in continuous-variable systems. This provides an important supply  on generation of entangled states in discrete-variable systems \cite{Bennett1, Bennett2, Bennett3, Horodecki1, Beigi}.

(ii) Theorem 4.1 tells us that resultant state remains genuine correlation as long as all source states are genuinely correlative.
This implies that multiple choices of the set of source states $\{|\phi_i\rangle\}$ are realistic for creating a multimode Gaussian state with certain genuine correlation. This information is valuable in the situation when one is forced to prepare Gaussian states with lower mode in laboratory in order to generate multimode Gaussian states by our protocol.
 It is due to the fact that preparing source states like photos in some physical substrates is difficult.
Multiple choices also means there are multiple plans information distribution of quantum Gaussian networks. It is wellknown that design of information distribution between multiple nodes is a challenging problem in quantum domains yet \cite{Kimble}.
In fact, one can compute the mean value and the standard deviation of $\mathcal{ GM}$ corresponding to different source states. The design of lower mean and lower standard deviation mean lower cost on average and stronger stability of quantum networks. Thus the nonuniqueness of the set of source states
is also a crucial point of our protocol.

(iii) For any $0<c<1$, we can create an n-mode pure Gaussian state $\rho$ with $\mathcal{ GM}(\rho)=c$ from 2-mode pure Gaussian states and 3-mode pure three single-mode squeezed states in a tritter (see Section III). For example,  one can create a 7-mode pure Gaussian state $\rho$ with $\mathcal{ GM}(\rho)=c$  by applying two 2-mode Gaussian unitary operations over two 2-mode pure Gaussian states and one 3-mode pure three single-mode squeezed state in a tritter. The suitable parameter selection of such source states can guarantee that the resultant state $\rho$ satisfies the condition $\mathcal{ GM}(\rho)=c$.


By Theorem 4.1, one can see that another critical point in implementing our protocol is to find out the optimal Gaussian unitary operations $\{U_i\}$. Note that every 2-mode Gaussian unitary operation $U_i$ is determined by a  $4\times 4$ symplectic matrix $S_i$ (see Section II), we will provide a one-parameter classification of $S_i$ in order to identify the optimal Gaussian unitary operations. For fluency of paper, such one-parameter classification is placed in appendix.
Based on such one-parameter classification, the optimal Gaussian unitary operations $\{U_i\}$ can be given as follows.

{\bf Theorem 4.2.} If the CM of $|\phi_i\ra$ reads as $$\Gamma_{|\phi_i\ra}=\left(\begin{array}{ccc} \gamma_1^{(i)}I_2 &C_{12}^{(i)}&C_{13}^{(i)}\\(C_{12}^{(i)})^{t}& A^{(i)}& C_{23}^{(i)}\\ (C_{13}^{(i)})^{t}& (C_{23}^{(i)})^{t}&\gamma_2^{(i)}I_2\end{array}\right),$$ here $\gamma_1^{(i)}\geq 1,\gamma_2^{(i)}\geq 1,I_2$ is the $2\times 2$ unit matrix, $A^{(i)}$ is a $(2n_i-4)\times (2n_i-4)$ matrix, $|\phi_i\ra$ is $n_i-$mode, then the optimal Gaussian unitary operation $U_i$ can always be designed as Table I, here $\lambda_i$ is one-parameter classification of symplectic matrix $S_i$ determining  $U_i$.

\begin{widetext}\begin{center}

\begin{table}[htbp]
	\centering
\caption{ }
		\begin{threeparttable}\begin{tabular}{|c|c|}\hline

Type I  &  $\lambda_i^2\geq \frac{-(\gamma_2^{(i)}+\gamma_1^{(i+1)})^2+\sqrt{(\gamma_2^{(i)}+\gamma_1^{(i+1)})^4+4(\gamma_2^{(i)}+
            \gamma_1^{(i+1)})^2\gamma_2^{(i)}\gamma_1^{(i+1)}(\gamma_2
^{(i)}-1)}}
{2(\gamma_2^{(i)}+\gamma_1^{(i+1)})^2}$ \\ \hline
 Type II  &  $\lambda_i^2\geq \frac{(\gamma_2^{(i)}+\gamma_1^{(i+1)})^2+\sqrt{(\gamma_2^{(i)}+\gamma_1^{(i+1)})^4+4(\gamma_2^{(i)}+\gamma_1^{(i+1)})^2\gamma_2^{(i)}
                  \gamma_1^{(i+1)}(\gamma_2^{(i)}-1)}}{2(\gamma_2^{(i)}+\gamma_1^{(i+1)})^2}$ \\ \hline
Type III &  $\lambda_i^2\geq \frac{-(\gamma_2^{(i)}+\gamma_1^{(i+1)})^2+\sqrt{((\gamma_2^{(i)})^2+(\gamma_1^{(i+1)})^2)^2+4\gamma_2^{(i)}\gamma_1^{(i+1)}((\gamma_2^{(i)})^2-1)}}
{2\gamma_2^{(i)}\gamma_1^{(i+1)}}$  \\   \hline
 Type IV& $(\gamma_2^{(i)}\lambda_{i1}^2+\gamma_1^{(i+1)})(\gamma_1^{(i+1)}\lambda_{i2}^2+\gamma_2^{(i)})\geq (\gamma_2^{(i)})^3\gamma_1^{(i+1)}$ \\
 \hline
\end{tabular}
\end{threeparttable} \end{table}\end{center}\end{widetext}

To identify optimal Gaussian unitary operations by Theorem 4.2, we consider a simple scenario of a chain or a star network consisting of three  identical  three single-mode squeezed states in a tritter $|\phi_{\gamma}\ra$ (Fig. 2).
In Section III, it is shown  $$\mathcal {GM}(|\phi_{\gamma}\ra)=1-\frac {81}{(5+4\cosh(4\gamma))^2}.$$ In the case of a chain network (Fig. 2(a)), we apply $2$-mode Gaussian unitary operation $U_1$, $U_2$ on party $P_2$ and $P_3$ respectively.
The resultant state denoted by $|\psi_{U_1,U_2}\ra$  is a $9-$mode state,
$$\begin{array}{ll}
  & |\psi_{U_1,U_2}\ra
= \Phi(\otimes^3|\phi_{\gamma}\ra) \\
 = & I_1\otimes U_1\otimes U_2\otimes I_4\otimes I_5\otimes I_6\otimes I_7(\otimes^3|\phi_{\gamma}\ra)).\end{array}$$
 In the case of a star network (Fig. 2(b)), we apply $2-$mode unitary $U_1$ on $P_{51}$ and $P_{52}$, $U_2$ on $P_{51}$ and $P_{53}$ respectively. $\Phi_1$, $\Phi_2$ is defined as following:
$$\Phi_1(\otimes^3|\phi_{\gamma} \rangle)=(\otimes _{i=1}^4I_i \otimes U_1 \otimes I_{53}\otimes I_6\otimes I_7)(\otimes^3|\phi_{\gamma} \rangle) ,$$
$$\Phi_2\Phi_1(\otimes^3|\phi_{\gamma} \rangle)=(\otimes _{i=1}^4I_i \otimes I_{52}\otimes U_2\otimes I_6\otimes I_7)(\Phi_1(\otimes^3|\phi_{\gamma} \rangle)) .$$
The resultant state $|\psi_{U_1,U_2}\ra= \Phi_2\Phi_1(\otimes^3|\phi_{\gamma} \rangle)$.
 Theorem 4.1 tells that
$$\max_{U_1,U_2}\mathcal{GM}(|\psi_{U_1,U_2}\ra)=\mathcal {GM}(|\phi_{\gamma}\ra)=1-\frac {81}{(5+4\cosh(4\gamma))^2}.$$
If the maximum is reached at some $U_1,U_2$, we say $U_1,U_2$ are optimal Gaussian unitary operations.
We will find optimal Gaussian unitary operations of type I in Theorem 4.2.
Note that there is a local Gaussian unitary $U$ such that $U|\phi_{\gamma}\ra$ has the CM
$$\Gamma_{|\phi_{\gamma}\ra}=\left(\begin{array}{ccc} \sqrt{\mathcal{R}_+\mathcal{R}_- } I&C_{12}&C_{13}\\C_{12}^{t}& C_{22}& C_{23}\\ C_{13}^{t}& C_{23}^{t}&\sqrt{\mathcal{R}_+\mathcal{R}_- }I\end{array}\right).$$
The table I of Theorem 4.2 shows that symplectic matrices of type I determining $U_i(i=1,2)$ satisfy the condition
$$\lambda^2\geq \frac{\sqrt {\mathcal{R}_+\mathcal{R}_- }-1}{2}.$$
Hence  $U_1,U_2$ are $2-$mode squeezing operation. Recall that a $2-$mode squeezing operation is  an active transformation which models the physics of optical parametric amplifiers and is routine to create CV entanglement. It acts on the pair of modes $i$  and $j$ via the unitary
$$\hat{U}_{i,j}(\xi)=\exp[\xi(\hat{a}_i^{\dag}\hat{a}_j^{\dag}-\hat{a}_i \hat{a}_j )].$$ Furthermore, it corresponds to the symplectic matrices of type I  with $\lambda=\cosh \xi$ \cite{FAdesso}.  Thus $2-$mode squeezing operations with $\cosh^2 \xi\geq \frac {\sqrt{5+4\cosh(4\gamma)}-3}{6}$ are  optimal Gaussian unitary operations. Additionally, the table I of Theorem 4.2 also provides some other possible choices of optimal Gaussian unitary operations.

\begin{figure}[htbp]
\centering
\includegraphics[scale=0.4]{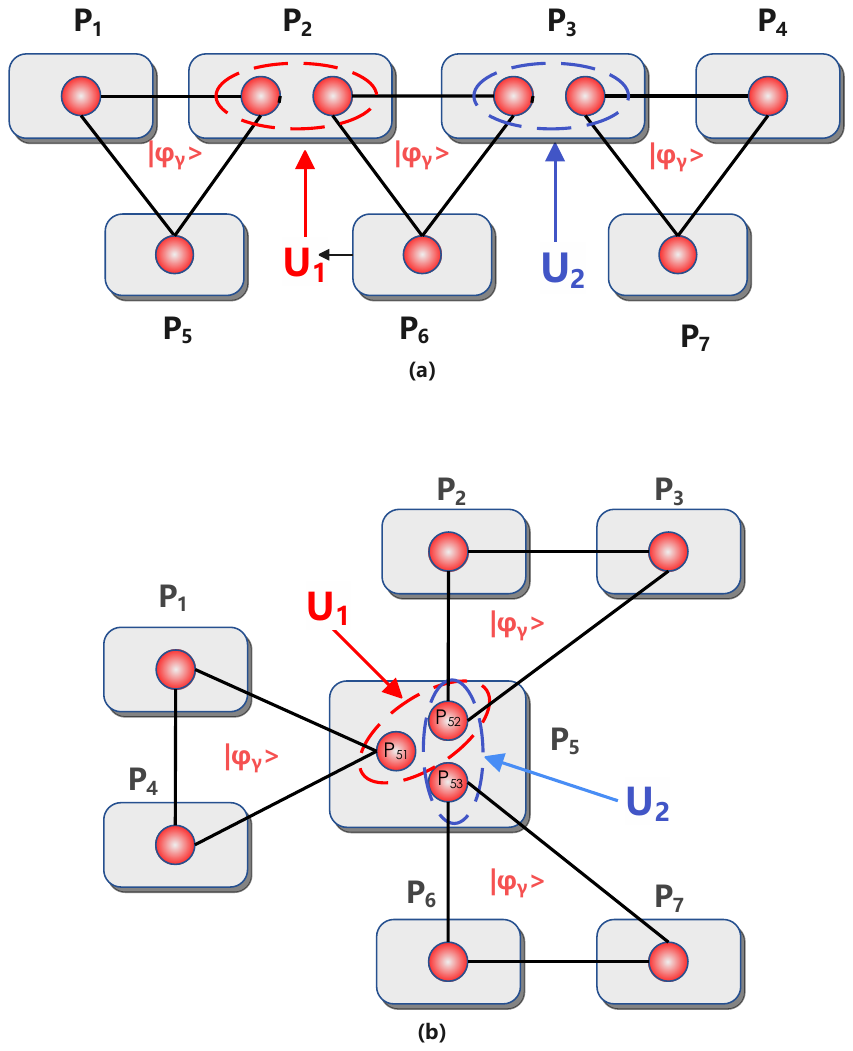}\caption{Protocols in a chain or a star network with three identical $3-$mode squeezed vacuum states as source states.}
\end{figure}

\vspace{0.1in}


\section{Conclusion}

Gaussian networks are fundamental in network information theory. Here senders and receivers are connected through diverse routes
that extend across intermediate sender-receiver pairs that act as nodes. The quantum network is Gaussian if the operations at the nodes
and the final state shared by end-users are Gaussian. Although classical Gaussian networks is established rigorously, the quantum analogue is far from mature \cite{Ghalaii}. Therefore, it is interesting to find the Gaussian network mechanism for creating a multimode state having certain amount of genuine correlation.

In this paper, we present a deterministic scheme for generating Gaussian states with certain amount of GGQC and distribute them in the form of Gaussian quantum networks. Given limited amount of sources, our scheme can generate genuinely correlative Gaussian states (including genuinely entangled Gaussian states) with the application of optimal Gaussian unitary operations. An explicit description of optimal Gaussian unitary operations is also provided.

Our choice for generating Gaussian states with certain amount of GGQC is not unique since there are multiple choices of optimal Gaussian unitary operations and source states. This raises naturally one interesting question  whether all these choices are
equivalent, or a subset of these choices are more beneficial. It is key to comprehend the mechanism of information distribution in quantum Gaussian networks \cite{Kimble}.

\vspace{0.1in}
{\it Acknowledgement}

We thank professor Jinchuan Hou for helpful discussion. We acknowledge that the research was  supported by NSF of China (12271452,11671332) and NSF of
Fujian (2023J01028).
\vspace{1mm}

{\it Data availability statement}

All data that support the findings of this study are included within the article.




\vspace{1mm}

{\it Additional Information}

Correspondence  should be addressed to Shuanping Du.

\vspace{1mm}

{\it\bf Appendix : Proof of our results}\vspace{0.1in}

In order to state optimal Gaussian unitary operations clearly, we need
 to  classify 2-mode symplectic matrices.

 \textbf{Proposition 1.}
{\it For $S\in {\rm Sp}(4,\mathbb R)$, there are $L,R\in {\rm Sp}(4,\mathbb R)$ with the form $L=L_1\oplus L_2$, $R=R_1\oplus R_2$ such that $LSR$ has the one of the following forms:\\
{\tiny $S_I= \left(\begin{array}{cccc}\sqrt{\lambda^2+1} &0 & \lambda &0\\
                               0&\sqrt{\lambda^2+1} & 0& -\lambda\\
\lambda&0&\sqrt{\lambda^2+1} &0 \\
                              0&-\lambda&0 &\sqrt{\lambda^2+1} \end{array}\right)$, $\lambda> 0$;\\
 $S_{II}=\left(\begin{array}{cccc} \sqrt{\lambda^2-1}&0 &\lambda &0\\
                               0& -\sqrt{\lambda^2-1}& 0&\lambda\\
                               \lambda  &0&\sqrt{\lambda^2-1}&0 \\
                              0&\lambda&0 &-\sqrt{\lambda^2-1} \end{array}\right)$, $\lambda>1$;\\
 $S_{III}=\left(\begin{array}{cccc}1 &0 & \lambda &0\\
                               0&1 & 0& 0\\
                              0&0&1 &0 \\
                              0&-\lambda&0 &1 \end{array}\right)$, $\lambda\in \mathbb{R}$;\\
 $S_{IV}=\left(\begin{array}{cccc}\lambda_1 &0 & 1&0\\
                               0&0 &0&1\\
                              1&0&0 &0 \\
                              0&1&0 &\lambda_2  \end{array}\right)$, $\lambda_1,\lambda_2\in \mathbb{R}$;\ \ \ \ \ \ \
$S_{V}=\left(\begin{array}{cccc}0 &0 & 1 &0\\
                               0&0 & 0& 1\\
                              -1&0&0 &0 \\
                              0&-1&0 &0 \end{array}\right)$.\\
 $S_{VI}=\left(\begin{array}{cccc}\sqrt{1-\lambda^2} &0 & \lambda &0\\
                               0&\sqrt{1-\lambda^2} & 0& \lambda\\
                              -\lambda&0&\sqrt{1-\lambda^2} &0 \\
                              0&-\lambda&0 &\sqrt{1-\lambda^2} \end{array}\right)$,  $0<\lambda< 1$;\\}}


{\bf Proof of Proposition 1.} Write $S=\left(\begin{array}{cc}
                             S_{11}&S_{12}\\
                             S_{21}&S_{22}\end{array}\right)$.
A direct computation shows that $S\in {\rm Sp}(4,\mathbb R)$ if and only if the following hold true:
\begin{equation}\left\{ \begin{array}{l}
\det(S_{11})+\det(S_{12})=1\\
\det(S_{21})+\det(S_{22})=1\\
S_{11}\Delta S_{21}^{t}+S_{12}\Delta S_{22}^{t}=0.\end{array}\right.\label{a1}\end{equation}
Moreover
\begin{equation}\left\{ \begin{array}{l}
\det(S_{11})=\det(S_{22})\\
 \det(S_{12})=\det(S_{21}).
 \end{array}\right.\label{a2}\end{equation}
From the singular value decomposition, $L_{i1}S_{ii}R_{i1}=\text{ diag}(\lambda_{i1},\lambda_{i2})$, here $\det(L_{i1})=\det(R_{i1})=1$, $\lambda_{i1},\lambda_{i2}\in \mathbb R$, $\lambda_{i1}\geq0$,  $\lambda_{i1}\lambda_{i2}=\det(S_{11})$, $i=1,2$. Take $$L_1=\left(\begin{array}{cc} L_{11}&0\\
                       0&L_{21}\end{array}\right), \ \ R_1=\left(\begin{array}{cc} R_{11}&0\\
                     0&R_{21}\end{array}\right),$$  we have
\begin{equation}L_1SR_1=\left( \begin{array}{cc}\text{ diag}( \lambda_{11},\lambda_{12} )& S_{12}'\\
                                                              S_{21}' & \text{ diag}( \lambda_{21},\lambda_{22} )\\
                                                              \end{array}\right)=S'.\label{a3}\end{equation}
Next, we  divide four cases according to the value of $\det(S_{11})$.

{\bf Case 1.} $\det(S_{11})\neq 0 \text{ or } 1$.

$L_{2}=\text {diag}(\sqrt{|\frac {\lambda_{12}}{\lambda_{11}}}|,\sqrt{|\frac {\lambda_{11}}{\lambda_{12}}|},\sqrt{|\frac {\lambda_{22}}{\lambda_{21}}|},\sqrt{|\frac {\lambda_{21}}{\lambda_{22}}|})$.
Then $L_2L_1SR_1$ has the form
\begin{equation}\left( \begin{array}{cc} \sqrt{|\lambda_{11}\lambda_{12}|}\text{ diag}(1, \frac{\lambda_{12}}{|\lambda_{12}|} )& S_{12}''\\
                                                              S_{21}'' & \sqrt{|\lambda_{21}\lambda_{22}|}\text{ diag}(1,\frac{\lambda_{22}}{|\lambda_{22}|} )\end{array}\right),\label{a4}\end{equation} denoted by $S''$.
Applying the singular value decomposition to $S_{12}''$, we have unitary $U,V$ such  that  $U\text{ diag}(1, \frac{\lambda_{12}}{|\lambda_{12}|} )S_{12}''V\text{ diag}(1,\frac{\lambda_{22}}{|\lambda_{22}|} )=\text{diag} (\beta_1,\beta_2)$, here $U,V$ are $2\times 2$ real unitaries and $\det(U)=\det(V)=1$, $\beta_1\beta_2=\det(S_{12}'')$, $\beta_1>0$. Take
\begin{equation}\left\{\begin{array}{l}
L_3=\text{diag}(U\text{ diag}(1, \frac{\lambda_{12}}{|\lambda_{12}|} ),V^{t}\text{ diag}(1,\frac{\lambda_{22}}{|\lambda_{22}|} )),\\
R_2=\text{diag}(U^{t}\text{ diag}(1, \frac{\lambda_{12}}{|\lambda_{12}|} ),V\text{ diag}(1,\frac{\lambda_{22}}{|\lambda_{22}|} )),\\
L_4=\text{diag}(\sqrt[4]{|\frac{\beta_2}{\beta_1}|}, \sqrt[4]{|\frac{\beta_1}{\beta_2}|},\sqrt[4]{|\frac{\beta_2}{\beta_1}|},\sqrt[4]{|\frac{\beta_1}{\beta_2}|}),\\
R_3=\text{diag}(\sqrt[4]{|\frac{\beta_1}{\beta_2}|}, \sqrt[4]{|\frac{\beta_2}{\beta_1}|},\sqrt[4]{|\frac{\beta_1}{\beta_2}|},\sqrt[4]{|\frac{\beta_2}{\beta_1}|}).
\end{array}\right.\label{a5}\end{equation}
It can be checked that $L_4L_3S''R_2R_3$ has the form $$\left( \begin{array}{cc} \sqrt{\lambda_{11}\lambda_{12}}\text{ diag}(1, \frac{\lambda_{12}}{|\lambda_{12}|} ) &\sqrt{|\beta_1\beta_2|}\text{diag}(1,\frac{\beta_2}{|\beta_2|})\\ S_{21}''' &\sqrt{\lambda_{21}\lambda_{22}}\text{ diag}(1,\frac{\lambda_{22}}{|\lambda_{22}|} )\end{array}\right),$$ denoted by $S'''$.
Take $\lambda=\sqrt{|\beta_1\beta_2|}$. From Equations (4) and (5), it follows that $S'''$ has the form $S_I$, $S_{II}$, $S_{VI}$ according to $\det(S_{11})>1$, $0<\det(S_{11})<1$, and $\det(S_{11})<0$, respectively.

{\bf Case 2.} $\det(S_{11})= 1$.

In this case, $\det(S_{12})= 0$ and $\beta_2=0$. Following the Equation \eqref{a4}, and taking $U,V$ as in Case I, we choose $L_5=\text{diag}(U,V^{t})$, $R_4=\text{diag}(U^{t},V)$, $\lambda=\beta_1$ and obtain that $L_5L_2L_1SR_1R_4$ has the form $S_{III}$.

{\bf Case 3.} $\det(S_{11})= 0$ and $S_{11}\neq 0$.

In this case, $\lambda_{i2}=0$. From Equation \eqref{a3}, it follows that $$L_1SR_1=\left( \begin{array}{cc}\text{ diag}(
                                                             \lambda_{11},0 )& S_{12}'\\
                                                              S_{21}' & \text{ diag}( \lambda_{21},0 )\\
                                                              \end{array}\right)=S'.$$
Write $S_{12}'=\left( \begin{array}{cc} x_{11} & x_{12}\\ x_{21} & x_{22}\end{array}\right)$ and $S_{21}'=\left( \begin{array}{cc} y_{11} & y_{12}\\ y_{21} & y_{22}\end{array}\right)$. Substituting them into Equation \eqref{a2}, we obtain $ x_{22} =y_{22}=0$. Moreover, $x_{12}x_{21}=x_{12}x_{21}=-1$.
Let $$L_6=\text{diag}(\frac{x_{12}}{|x_{12}|}\sqrt[4]{|\frac{x_{21}}{x_{12}}|},-\frac{x_{21}}{|x_{21}|}\sqrt[4]{|\frac{x_{12}}{x_{21}}|},
\sqrt[4]{|\frac{x_{21}}{x_{12}}|},\sqrt[4]{|\frac{x_{12}}{x_{21}}|}),$$
$$R_5=\text{diag}(\frac{x_{12}}{|x_{12}|}\sqrt[4]{|\frac{x_{12}}{x_{21}}|},-\frac{x_{21}}{|x_{21}|}\sqrt[4]{|\frac{x_{21}}{x_{12}}|},
\sqrt[4]{|\frac{x_{12}}{x_{21}}|},\sqrt[4]{|\frac{x_{21}}{x_{12}}|}),$$
It is checked that $$L_6S'R_5=\left(\begin{array}{cccc} \lambda_{11} &0 &x_{11}'&1 \\
                                                       0 &0 &-1&0\\
                                                       y_{11}'& y_{12}'&\lambda_{21} &0\\
                                                       y_{21}'& 0 &0 &0\end{array}\right)=S''.$$
Now let $L_7=\left(\begin{array}{cc}1&x_{11}'\\ 0&1\end{array}\right)\oplus \left(\begin{array}{cc}1&-\frac{y_{11}'}{y_{21}'}\\ 0&1\end{array}\right)$. We have $$L_7S''= \left(\begin{array}{cccc} \lambda_{11} &0 &0&1 \\
                                                       0 &0 &-1&0\\
                                                       0&  y_{12}'&\lambda_{21} &0\\
                                                        y_{21}'& 0'&0 &0\end{array}\right)=S'''.$$
Take  $$L_8=\left(\begin{array}{cccc} 1 &0&0&0\\
                                       0&1&0&0\\
                                     0& 0& 0& -y_{12}'\\
                                     0&0& -y_{21}'&0 \end{array}\right),$$ $$R_6=I\oplus \left(\begin{array}{cc} 0&-1\\1&0 \end{array}\right).$$ From Equation \eqref{a1}, $ y_{12}'y_{21}'=-1$ and $L_8$ is symplectic.
Now it can be  checked directly that  that  $L_8S'''R_6$ has the form $S_{IV}$.

{\bf Case 4.} $S_{11}=0$.

From Equations (4) and (5), one gets $S_{22}=0$, $\det({S_{12}})=\det(S_{21})=1$. Applying the singular value decomposition to $S_{12}$ and $S_{21}$, we can find $U_i,V_i$ such that $U_1S_{12}V_1=\text{diag} (\beta_1,\beta_2)$, $U_2S_{21}V_2=\text{diag} (\beta_3,\beta_4)$,  $U_i,V_i$ are $2\times 2$ real unitaries and $\det(U_i)=\det(V_i)=1$, $\beta_i\rangle0$ ($i=1,\ldots, 4$).
Take $$L_9=\text{diag}(\sqrt{\frac{\beta_2}{\beta_1}}, \sqrt{\frac{\beta_1}{\beta_2}})U_1\oplus \text{diag}(\sqrt{\frac{\beta_4}{\beta_3}}, \sqrt{\frac{\beta_3}{\beta_4}})U_2,$$ $$R_7=V_2\oplus V_1.$$
Then $L_9SR_7$ has the form $S_{V}$.

\vspace{0.1in} To prove Theorem 4.1 and Theorem 4.2, we consider a simple  scenario having
three parties and two sources [see Fig.3]. Here the first two parties
share a $m$-mode state $\rho_{1}$, the second and third parties share a $n$-mode state
$\rho_{2}$, and the central party is performed 2-mode Gaussian unitary operations. The resultant $(m+n)$-mode state reads $$\sigma=(I\otimes U\otimes I)(\rho_1\otimes \rho_2)(I\otimes U\otimes I)^{\dag}.$$
\begin{figure}[htbp]
\centering
\includegraphics[scale=0.4]{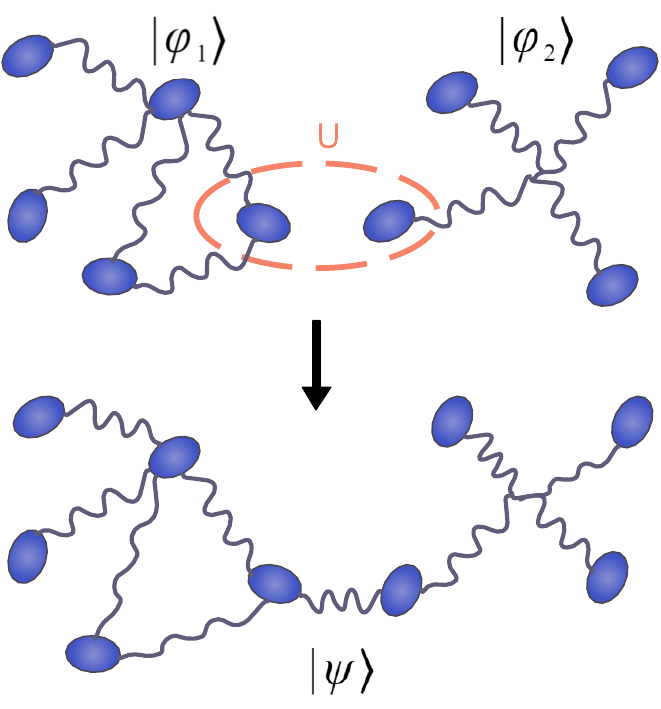}
\caption{ Schematic representation of 2-mode Gaussian unitary operations acting on two sources.}
\end{figure}

The general protocol can be reduced to this  scenario. Taking the example in Fig.1 again,
$$\begin{array}{l}\Phi_1(\rho)=|\psi_1\rangle \otimes_{i=3}^6|\phi_i\rangle,\\
 \Phi_2\Phi_1(\rho)=|\psi_2\rangle\otimes _{i=4}^6|\phi_i\rangle,\\
\Phi_3\Phi_2\Phi_1(\rho)=|\psi_3\rangle\otimes _{i=5}^6|\phi_i\rangle,\\
 \prod_{i=4}^1\Phi_i(\rho)=|\psi_4\rangle\otimes |\phi_6\rangle,\end{array}$$
 here $$\begin{array}{l}|\psi_1\rangle=(I_1\otimes I_2\otimes U_1\otimes I_{34}\otimes I_{ 41}\otimes I_{42})(|\phi_1\rangle\otimes |\phi_2\rangle),\\
 |\psi_2\rangle=(I_1\otimes I_2\otimes I_{31}\otimes I_{33}\otimes U_2\otimes I_{ 41}\otimes I_{42}\otimes I_{51})\\ \ \ \ \ \ \ \ \ (|\psi_1\rangle\otimes |\phi_3\rangle),\\
 |\psi_3\rangle=((\otimes_{i=1}^4 I_i)\otimes U_3\otimes I_{71})(|\psi_3\ra\otimes |\phi_4\ra)\\ \ \ \ \ \ \ \ \ (|\psi_2\rangle\otimes |\phi_4\rangle),\\
 |\psi_4\rangle=((\otimes_{i=1}^6 I_i)\otimes U_4\otimes I_{ 73}\otimes I_{8})(|\psi_3\ra\otimes |\phi_5\ra). \end{array}$$
 If Theorem 4.1 holds true in the simple scenario, then
 $$\begin{array}{l}
 \max_{U_5} \mathcal{ GM}(\Phi(\rho))=\min \{\mathcal{ GM}(|\psi\ra_4),\mathcal{ GM}(|\phi\ra_6)\},\\
 \max_{U_4}\mathcal{ GM}(|\psi_4\rangle)= \min \{\mathcal{ GM}(|\psi\ra_3),\mathcal{ GM}(|\phi\ra_5)\},\\
 \cdots\\
 \max_{U_1}\mathcal{ GM}(|\psi\ra_1) = \min \{\mathcal{ GM}(|\phi\ra_1),\mathcal{ GM}(|\phi\ra_2)\}.\end{array}$$
 Thus $\max_{\{U_i\}} \mathcal{ GM}(\Phi(\rho))=\min_i \{\mathcal{ GM}(|\phi\ra_i)\}$.

{\bf Proof of Theorem 4.1.} For any $(m+n)$-mode initial Gaussian state $\rho=|\phi_1\ra\otimes | \phi_2\ra$, under some suitable local Gaussian unitary
operation, its CM can be reduced to the
 form $$\Gamma_{\rho}=\left(\begin{array}{cc}\begin{array}{cc} A &C\\C^{t}&\gamma_1I_2\end{array} & 0\\ 0& \begin{array}{cc} \gamma_2  I_2 &D\\D^{t}&B\end{array}\end{array}\right),$$ where
 $A=(A_{ij})_{(m-1)\times (m-1)}$, $B=(B_{ij})_{(n-1)\times (n-1)}$, $C=(C_{1m}, C_{2m}, \dots, C_{(m-1)m})^{t}$, $D=(D_{12}, D_{13}, \dots, D_{1n})$, $A_{ij},B_{ij},C_{ij},D_{ij}\in {\mathcal M}_2(\mathbb R)$.

 Firstly, we show that  $$\mathcal{ GM}(\sigma_{U})\leq \min \{\mathcal{ GM}(|\phi\ra_1),\mathcal{ GM}(|\phi\ra_2)\}.$$ Assume that $$\mathcal{ GM}(|\phi_1\ra)=\mathcal M_{\phi_1}(\alpha_1),$$ $$\mathcal{ GM}(|\phi_2\ra)=\mathcal M_{\phi_2}(\alpha_2),$$
 here $\alpha_1\subset \{1,\ldots, m\}$ and $\alpha_2\subset \{1,\ldots, n\}$.
If $S=\left(\begin{array}{cc}
                             S_{11}&S_{12}\\
                             S_{21}&S_{22}\end{array}\right)$, then the resultant state $\sigma_U$ has the CM $\Gamma_{\sigma_U}$
{\tiny $$\left( \begin{array}{cccc} A  & CS_{11}^{t} & CS_{21}^{t} &0\\
 S_{11}C^{t}  &  \gamma_1S_{11}S_{11}^{t}+\gamma_2S_{12}S_{12}^{t} & \gamma_1S_{11}S_{21}^{t}+\gamma_2S_{12}S_{22}^{t}
                           & S_{12}D\\
 S_{21}C^{t} &\gamma_1S_{21}S_{11}^{t}+\gamma_2S_{22}S_{12}^{t} &\gamma_1S_{21}S_{21}^{t}+\gamma_2S_{22}S_{22}^{t}
                          & S_{22}D\\
0& D^{t}S_{12}^{t} & D^t S_{22}^{t} &B
\end{array}    \right).$$}
Without loss of generality, we may assume $m\notin \alpha_1$. Otherwise, we just replace $\alpha_1$ with the complement set of $\alpha_1$ . A direct computation shows that
$$\begin{array}{ll}
&{\mathcal M}_{\sigma_U}(\alpha_1)\\
= &1-\frac 1 {\mathcal D_{\sigma_U}(\alpha_1)\mathcal D_{\sigma_U}(\{1,\cdots,m+n\}\backslash \alpha_1) }\\
=& 1-\frac 1 {\mathcal D_{\rho}(\alpha_1)\mathcal D_{\rho}(\{1,\cdots,m+n\}\backslash \alpha_1) }\\
=&1-\frac 1 {\mathcal D_{|\phi_1\ra}(\alpha_1)\mathcal D_{|\phi_1\ra}(\{1,\cdots,m\}\backslash \alpha_1)}\\
=&\mathcal{ GM}(|\phi_1\ra),\end{array}$$
$$\begin{array}{ll}
&{\mathcal M}_{\sigma_U}(m+\alpha_2)\\
=& 1-\frac 1 {\mathcal D_{\sigma_U}(m+\alpha_2)\mathcal D_{\sigma_U}(\{1,\cdots,m+n\}\backslash (m+\alpha_2))}\\
 =  &1-\frac 1 {\mathcal D_{\rho}(m+\alpha_2)\mathcal D_{\rho}(\{1,\cdots,m+n\}\backslash (m+\alpha_2))}\\
 = & 1-\frac 1 {\mathcal D_{|\phi_2\ra}(\alpha_2)\mathcal D_{|\phi_2\ra}(\{1,\cdots,n\}\backslash \alpha_2)}\\
 = &\mathcal{ GM}(|\phi_2\ra),\end{array}$$
here $m+\alpha_2=\{m+s\mid s\in\alpha_2\}$. Therefore $\mathcal{ GM}(\sigma_{U})\leq \min \{\mathcal{ GM}(|\phi\ra_1),\mathcal{ GM}(|\phi\ra_2)\}$.
This deduces  $\max_{\{U_i\}}\mathcal{ GM}(|\psi\ra)\leq\min_i \{\mathcal{ GM}(|\phi\ra_1, |\phi\ra_2)\}.$

Next, we show that $$\mathcal{ GM}(\sigma_{U})=\min \{\mathcal{ GM}(|\phi\ra_1),\mathcal{ GM}(|\phi\ra_2)\}$$ for some $U$. By the definition, one only need verify, for every 2-partition $\mathcal M_{\sigma_{U}}\geq \min \{\mathcal{ GM}(|\phi\ra_1),\mathcal{ GM}(|\phi\ra_2)\}$.  Let $\alpha$ denote a subset of $\{1,\cdots, m+n\}$ and $\overline{\alpha}$ denote its complement set.
We have four cases:
 $\{m,m+1\}\subset \alpha$; $m\in\alpha$, $m+1\in \overline{\alpha}$;  $\{m,m+1\}\subset \overline{\alpha}$ ;
 $m+1\in\alpha$, $m\in \overline{\alpha}$.
Note that $\mathcal{ M}(\alpha)=\mathcal{ M}(\overline{\alpha})$, we only need treat the following cases.

Case 1 . $\{m,m+1\}\subset \alpha$, $\{m+2,\cdots, m+n\}\cap \overline{\alpha}\neq \emptyset$.

$$\begin{array}{ll}&\mathcal{ M}_{\sigma_U}(\alpha)=1-\frac 1 {\mathcal D_{\sigma_U}(\alpha)\mathcal D_{\sigma_U}(\overline{\alpha})}\\
                  \geq  & 1-\frac {\mathcal D_{\sigma_U}(\{m,m+1,\cdots, m+n\})} {\mathcal D_{\sigma_U}(\alpha\cap \{m,m+1,\cdots, m+n\})\mathcal D_{\sigma_U}(\overline{\alpha}\cap \{m,m+1,\cdots, m+n\})}  \\
                  =&   1-\frac {\mathcal D_{\rho}(\{m,m+1,\cdots, m+n\})} {\mathcal D_{\rho}(\alpha\cap \{m,m+1,\cdots, m+n\})\mathcal D_{\rho}(\overline{\alpha}\cap \{m,m+1,\cdots, m+n\})}\\
                   =&   1-\frac {\gamma_1^2} {\gamma_1^2\mathcal D_{\rho}(\alpha\cap \{m+1,\cdots, m+n\})\mathcal D_{\rho}(\overline{\alpha}\cap \{m+1,\cdots, m+n\})}\\
                   =& \mathcal{ M}_{|\phi_2\ra}(\alpha\cap \{m+1,\cdots, m+n\}).
                  \end{array}  $$
 The inequality follows from the correlation ${\mathcal M(\rho)}$ is nonincreasing under kickout \cite{Hou}.

Case 2. $\{m,m+1\}\subset \alpha$, $\{1,\cdots, m-1\}\cap\overline{\alpha}\neq \emptyset$.

$$\begin{array}{ll}&\mathcal{ M}_{\sigma_U}(\alpha)=1-\frac 1 {\mathcal D_{\sigma_U}(\alpha)\mathcal D_{\sigma_U}(\overline{\alpha})}\\
                  \geq  & 1-\frac {\mathcal D_{\sigma_U}(\{1,2,\cdots,m, m+1\})} {\mathcal D_{\sigma_U}(\alpha\cap \{1,2,\cdots,m, m+1\})\mathcal D_{\sigma_U}(\overline{\alpha}\cap \{1,2,\cdots,m, m+1\})}  \\
                  =&   1-\frac {\mathcal D_{\rho}(\{1,2,\cdots,m, m+1\})} {\mathcal D_{\rho}(\alpha\cap \{1,2,\cdots,m, m+1\})\mathcal D_{\rho}(\overline{\alpha}\cap \{1,2,\cdots,m, m+1\})}\\
                   =&   1-\frac {\gamma_2^2} {\gamma_2^2\mathcal D_{\rho}(\alpha\cap \{1,2,\cdots,m\})\mathcal D_{\rho}(\overline{\alpha}\cap \{1,\cdots, m\})}\\
                   =& \mathcal{ M}_{|\phi_1\ra}(\alpha\cap \{1,2,\cdots,m\}).
                  \end{array}  $$

Case 3.  $m\in\alpha$, $m+1\in \overline{\alpha}$.

 $$\begin{array}{ll}&\mathcal{ M}_{\sigma_U}(\alpha)=1-\frac 1 {\mathcal D_{\sigma_U}(\alpha)\mathcal D_{\sigma_U}(\overline{\alpha})}\\
                  \geq  & 1-\frac {\mathcal D_{\sigma_U}(\{1,2,\cdots,m, m+1\})} {\mathcal D_{\sigma_U}(\alpha\cap \{1,2,\cdots,m, m+1\})\mathcal D_{\sigma_U}(\overline{\alpha}\cap \{1,2,\cdots,m, m+1\})}  \\
                   =&   1-\frac {\gamma_2^2} {\mathcal D_{\sigma_U}(\alpha\cap \{1,2,\cdots,m\})\mathcal D_{\sigma_U}(\overline{\alpha}\cap \{1,2,\cdots,m-1, m+1\})}.
                  \end{array}  $$
Let $A_{\alpha}=(A_{ij})$, $C_{\alpha}=(C_{im})$, $i,j\in \alpha\cap \{1,2,\cdots,m-1\}$. Then
 $$\begin{array}{ll}
 & \mathcal D_{\sigma_U}(\alpha\cap \{1,2,\cdots,m\})\\
 =& \det\left(\begin{array}{cc}A_{\alpha} & C_{\alpha}S_{11}^{t}\\
    S_{11}C_{\alpha}^{t}& \gamma_1S_{11}S_{11}^{t}+\gamma_2S_{12}S_{12}^{t}\end{array}\right)\\
 = & \det(\gamma_1S_{11}S_{11}^{t}+\gamma_2S_{12}S_{12}^{t})\\
    &\det(A_{\alpha}-C_{\alpha}S_{11}^{t}(\gamma_1S_{11}S_{11}^{t}+
 \gamma_2S_{12}S_{12}^{t})^{-1}
                        S_{11}C_{\alpha}^{t}               )\\
 \overset{(*1)}\geq&\det(\gamma_1S_{11}S_{11}^{t}+\gamma_2S_{12}S_{12}^{t})\det(A_{\alpha}-C_{\alpha}C_{\alpha}^{t}/\gamma_1)\\
 =&\frac{\det(\gamma_1S_{11}S_{11}^{t}+\gamma_2S_{12}S_{12}^{t})\mathcal D_{|\phi_1\ra}(\alpha\cap \{1,2,\cdots,m\})}{\gamma_1^2},\end{array}$$
here the inequality ($*1$) is followed from \cite{Hou}.
Similarly,  let $A_{\overline{\alpha}}=(A_{ij})$, $C_{\overline{\alpha}}=(C_{im})$, $i,j\in \overline{\alpha}\cap \{1,2,\cdots,m-1\}$. Then
 $$\begin{array}{ll}
 & \mathcal D_{\sigma_U}(\overline{\alpha}\cap \{1,2,\cdots,m-1,m+1\})\\
 =& \det\left(\begin{array}{cc}A_{\overline{\alpha}} & C_{\overline{\alpha}}S_{21}^{t}\\S_{21}C_{\overline{\alpha}}^{t}& \gamma_1S_{21}S_{21}^{t}+\gamma_2S_{22}S_{22}^{t}\end{array}\right)\\
 = &\det(\gamma_1S_{21}S_{21}^{t}+\gamma_2S_{22}S_{22}^{t})\\
   &\det(A_{\overline{\alpha}}-C_{\overline{\alpha}}S_{21}^{t}
 (\gamma_1S_{21}S_{21}^{t}+\gamma_2S_{22}S_{22}^{t})^{-1}
                        S_{12}C_{\overline{\alpha}}^{t}               )\\
 \overset{(*2)}\geq&\det(\gamma_1S_{21}S_{21}^{t}+\gamma_2S_{22}S_{22}^{t})\det(A_{\overline{\alpha}}-C_{\overline{\alpha}}
 C_{\overline{\alpha}}^{t}/\gamma_1)\\
 =&\frac{\det(\gamma_1S_{21}S_{21}^{t}+\gamma_2S_{22}S_{22}^{t})\mathcal D_{|\phi_1\ra}(\overline{\alpha}\cap \{1,2,\cdots,m\})}{\gamma_1^2}.\end{array}$$
 the inequality ($*2$) is also from \cite{Hou}.
So  $$\begin{array}{ll}\mathcal{ M}_{\sigma_U}(\alpha)& \geq 1-  \frac{\mathcal D_{\sigma_U}(\{1,\ldots, m+1\})}{\mathcal D_{\sigma_U}(\alpha\cap \{1,\ldots, m\})\mathcal D_{\sigma_U}(\overline{\alpha}\cap\{1,\ldots,m-1, m+1\})}\\
&\geq 1-  \frac{A}{\mathcal D_{|\phi_1\ra}(\alpha)\mathcal D_{|\phi_1\ra}(\overline{\alpha})}\\
&\geq 1-  A\mathcal {M}_{|\phi_1\ra}(\alpha),\end{array}$$ here $$A=\frac{\gamma_1^4\gamma_2^2}{\det(\gamma_1S_{11}S_{11}^{t}+\gamma_2S_{12}S_{12}^{t})\det(\gamma_1S_{21}S_{21}^{t}+
\gamma_2S_{22}S_{22}^{t})}.$$
It is evident that we only need to find suitable $S=(S_{ij})$ such that
\begin{equation}\gamma_1^4\gamma_2^2\leq \det(\gamma_1S_{11}S_{11}^{t}+\gamma_2S_{12}S_{12}^{t})\det(\gamma_1S_{21}S_{21}^{t}+\gamma_2S_{22}S_{22}^{t}).\end{equation}
This will complete the proof of Theorem 4.1 and also find out optimal Gaussian unitary operations of our protocol.
 From Prop.4.2, we consider 6 types of $S$, respectively.
If $S$ is with Type I, then $S_{11}=S_{22}=\sqrt{\lambda^2+1}I_2$, $S_{12}=S_{21}=\text{diag}(\lambda,-\lambda)$,   for some $\lambda>0.$ A direct computation shows $$\begin{array}{ll}&\det(\gamma_1S_{11}S_{11}^{t}+\gamma_2S_{12}S_{12}^{t})\det(\gamma_1S_{21}S_{21}^{t}+\gamma_2S_{22}S_{22}^{t})\\
=&((\gamma_1+\gamma_2)\lambda^2+\gamma_1)^2
((\gamma_1+\gamma_2)\lambda^2+\gamma_2)^2.\end{array}$$ We can obtain that Eq.(9) is equivalent to $$\lambda^2\geq \frac{-(\gamma_1+\gamma_2)^2+\sqrt{(\gamma_1+\gamma_2)^4+4(\gamma_1+\gamma_2)^2\gamma_1\gamma_2(\gamma_1-1)}}{2(\gamma_1+\gamma_2)^2}.$$

Apply a similar process,  we can get  other  Types of $S$ (Table I).

\end{document}